\documentclass[a4paper,12pt]{article}
\usepackage[utf8]{inputenc}
\usepackage{graphicx}
\usepackage[top=1cm, bottom=2cm, left=2cm, right=1cm]{geometry}
\usepackage[dvipsnames]{xcolor}
\usepackage{cmap}
\usepackage{amsmath}
\usepackage{amssymb}
\graphicspath{{figures/}}
\DeclareGraphicsExtensions{.pdf,.png,.jpg,.png}

\def\prl{Phys. Rev. Lett.}

\usepackage{appendix}
 \usepackage{amsmath}
   \usepackage{amssymb}
 \usepackage{breqn}
 \usepackage{float}
 \graphicspath{{../img/}}
 \usepackage{multicol}
 \usepackage{verbatim}
 \usepackage{multirow}
 \usepackage{rotating}
 \usepackage{caption}
 \usepackage{subcaption}
 \usepackage{enumitem}
 \usepackage{hyperref}

\NewDocumentEnvironment{references}{}
{
    \begin{enumerate}{%
        \setlength{\itemsep}{4pt}%
        \setlength{\parskip}{0pt}%
        \setlength{\parsep}{0pt}%
        \setlength{\leftmargin}{0pt}%
    }
}
{
    \end{enumerate}
}

\NewDocumentCommand{\reference}{m}{%
    \item #1%
}
\begin{document}

\baselineskip 21pt

%opening
\title{\bf Analytical approaches for rapid prediction of gravitational waveforms for relativistic binary systems}

% }
\date{October 29, 2025}
\maketitle

% \author{\bf \hspace{-1.3cm}\copyright\, 2025 г. \ \
\centering A.V. Mishakina$^{1,2*}$, S.I. Blinnikov $^{1,3,4**}$\\
\begin{center}

$^1${\it Kurchatov Institute, Moscow, Russia}\\
% Институт теоретической и экспериментальной физики им. А.И. Алиханова, Москва, Россия}\\
$^2${\it MIPT, Moscow, Russia}\\
$^3${\it SAI MSU, Moscow, Russia}\\
$^4${\it VNIIA, Moscow, Russia}

\end{center}
\vspace{2mm}
%\received{~~~~~~~~}
% \received{ 00.00.2025 г. \\ После доработки 00.00.0000 г.; принята к публикации 00.00.0000 г.}
%\revised{}
\sloppypar
\vspace{2mm}
%\noindent

\abstract
We present a fast method for obtaining fully analytical approximations of gravitational waveforms from the merger of neutron stars and/or black holes during the earliest stages of the process.
The obtained analytical formula is compared with numerical calculations; its accuracy and limits of applicability are evaluated.
Our results may be useful not only for understanding the nature of binary systems in the earliest stages of evolution for gravitational-wave detectors but also for providing early alerts to gamma-ray, optical, and neutrino observatories.

{\it Keywords}: gravitational waves, gravitational waveforms, general theory of relativity, post-Newtonian formalism, matched filtering method

\vfill
\noindent\rule{8cm}{1pt}\\
{$^{*}$ sasha.mishakina@gmail.com}\\
{$^{**}$ sblinnikov@bk.ru}

\noindent
%%%%%%%%

\pagebreak
\section*{\centering{Introduction}}
\addcontentsline{toc}{section}{Introduction}

The gravitational-wave signal from a pair of merging black holes was first detected in 2015 by the detectors of LIGO observatory (Abbott et al., 2016).
Gravitational waves (GW) are of considerable interest, as they carry rich information about the parameters of merging black holes (BH) and neutron stars (NS)---their masses, spins, and orbital dynamics.
The analysis of GW makes it possible to test various relativistic theories of gravitation, as well as to study the formation of binary systems in the early Universe.

One of the most significant events was GW170817 gravitational-wave signal---the sixth detected gravitational-wave event and the first merger of the NS, accompanied by a short gamma-ray burst GRB170817A~(Abbott et al., 2017).
This event brilliantly confirmed the predictions of a short gamma-ray burst made in the papers of Blinnikov et al. (1984, 1990), however, it is still unclear what happened within about two seconds after the loss of the gravitational wave signal and before the registration of the gamma-ray burst.

Most GW detections are carried out using the matched filtering method (see, for example, Wainstein and Zubakov, 1960; Dhurandhar and Sathyaprakash, 1994).
A weak gravitational signal is detected in the noise by comparing it with one of the pre-modeled signals---a template.
Based on this, template banks are created (Owen, 1996; Ajith, et al., 2014; Allen, 2021; Coogan et al., 2022; Wadekar et al., 2024; Sharma et al., 2024).
The multidimensional space of parameters on which the waveform depends should be optimally covered by such patterns, minimizing computational costs while maintaining a high probability of detection.
In the book Dhurandhar, Mitra~(2022), in section 9.7.1, the difficulties of covering the parameter space with templates are described in detail.

The template banks used by LIGO are based either on the post-Newtonian (PN) formalism for the convergence phase, or on the results of numerical calculations in GR---Numerical Relativity (NR) for the merger and ringdown phases.
LIGO does not directly use numerical patterns in real time, but uses analytical approximations interpolated from numerical data (Cokelaer, 2007; Sakon et al., 2024).

However, in a number of cases, for example, at the initial search stage or in scenarios with a large number of variables and parameters, there is a need for faster analytical templates that would effectively identify GW signals from binary systems.
GW analytical templates provide high computing speed with low computational costs.

The purpose of this work is to obtain an analytical template for a close binary system at
the earliest stages of convergence using the computer algebra systems {\sc Maxima}
and {\sc Wolfram Mathematica}.
The motivation for us was the paper by Buskirk and Babiuc-Hamilton (2019), which claimed to
provide an entirely analytical solution, but actually used numerical solutions of
differential equations in {\sc Wolfram} language, which, of course, does not allow one to
calculate the GW-forms quickly.

In this note, we give only a brief description of the idea of our approach to a simplified
problem for spinless relativistic stars in a binary system with circular orbits in the
point mass approximation.
This approach is easily developed for the more general case of stars with spins and nonzero
orbital eccentricities.
Unlike Buskirk and Babiuc-Hamilton~(2019), we managed to obtain a fully analytical
expression.

\section{Post-Newtonian and other formalisms}
\label{subsec:PN}

According to the general theory of relativity, the evolution of a two-body system is
described by nonlinear Einstein equations that relate the curvature of space-time and the
distribution of energy and momentum (Landau and Lifshitz, 1988).
There are three stages in the evolution of a binary system: inspiral, merger, and ringdown.
The equations in the first and third stages can be reduced to a linear form and calculated
analytically, whereas for the pre-merger stage it is necessary to use NR methods.

At the inspiral stage, the binary system loses energy due to GW radiation.
Denote by $E(t)$ the dependence of the total energy of the binary system on time $t$, $\mathcal{F}$~ --- radiation loss power (Peters, 1964):
\begin{equation}
  \label{eq:eBalance}
  \frac{\mathrm{d}E}{\mathrm{d}t} = -\mathcal{F}.
\end{equation}
For example, in Tichy et al.~(2000), the following expression is proposed for $\mathcal{F}$ as a function of the velocity of circulation of $v$:
\begin{equation}
\label{F_V}
\mathcal{F}(v) =\frac{32 }{5}\eta^2  \frac{v^{10}}{c^{10}}
\left[1 + \sum_{k=2} f_k \frac{v^k}{c^k}
+ \sum_{k=6} g_k \ln \left(\frac{v}{c}\right) \frac{v^k}{c^k}  +\ldots\right] ,
\end{equation}
where $\eta = M_1 M_2/ M^2$ is the symmetric mass ratio in a binary system---see the definitions of these quantities below, after \eqref{rddot}, $c$ is the speed of light.
The ellipsis in \eqref{F_V} indicate possible terms proportional to the logarithms $(v/c)^k$ to various powers for $k\ge2$,
which may occur in higher orders of post-Newtonian corrections.
Tichy et al.~(2000) work in natural units of $G_N=c=1$.
Alternatively, an equation can be written for the loss of angular momentum by the system, but in the point mass approximation these two approaches are equivalent.

At the inspiral stage, the dynamics of the system is well described by the PN theory (Buonanno et al., 2009; Hannam et al., 2014; Taracchini et al., 2014).
The various orders of the post-Newtonian approximation are characterized by powers of a small parameter $x$, called the post-Newtonian parameter (PN parameter):
\begin{equation}
\label{PNparameter}
x=\frac{v^2}{c^2}.
\end{equation}
%{Здесь $v$ --- скорость обращения, $c$ --- скорость света.}

A correction of the order of $(v/c)^n$ to the Newtonian equation of motion is accounted for as an order of $n/2$ in the PN approximation.
% That is, for example, in 2.5PN order, the terms $(v/c)^5$ will appear.
Thus, the equation of motion for two bodies in the PN expansion is written symbolically as follows:
\begin{equation}
\ddot{\mathbf{r}} = -\frac{G_N M}{r^3} \mathbf{r} \bigg[1 + \sum_{n=2}^{\infty} \frac{A_{n/2}}{c^n }\bigg],
\label{rddot}
\end{equation}
where $r$ is the distance between the stars (in this paper we consider them point-like), $G_N$ is the Newtonian constant, $M = M_1+M_2$ is the total mass of the system of stars with masses of $M_1$ and $M_2$, $A_{n/2}$ --- expressions that depend on the physical parameters of the system (velocities, coordinates, masses, spins, etc.) and correspond to the $n/2$ order of PN approximation.

Below we use the coefficients from the formulas of the paper by Huerta et al. (2017), which is called the ``Complete waveform model for compact binaries on eccentric orbits''.
However, the full form here is not completely analytical, since, for example, the phase \(\Phi(t)\) requires numerical integration of equations for the evolution of the orbit.
Recently, even more long analytical expressions for the GW form have appeared, for example, the multipolar Postminkovsky (MPM) formalism developed by Blanchet, Damour and Iyer (Blanchet, 2024).
This set of approximation methods (called MPM-PN) has been successfully applied to compact binary systems, which made it possible to obtain equations of motion up to the fourth post-Newtonian order (4PN), and the shape of the GW and flow up to 4.5PN order, which surpasses the degree of approximation of the quadrupole Einstein formula (see also Dlapa et al., 2022; Dlapa et al., 2023).

The method proposed in this paper makes it possible to obtain a fully analytical expression for the evolution of the orbit in the various above-mentioned formalisms for the earliest stages of convergence, which are important for the rapid discovery of new sources of GW.

\section{{Transformation of expressions to analytical form}}
\label{sub:AnalytEquations}
Formula~\eqref{eq:eBalance} is rewritten in the form~(Buskirk, Babiuc-Hamilton, 2019) in terms of $x$ (see.~\ref{PNparameter}):
\begin{equation}
  \frac{\mathrm{d}x}{\mathrm{d}t} = -\frac{\mathcal{F}}{\mathrm{d}E/\mathrm{d}x}.
  \label{eq:dxdt}
\end{equation}

Approximants from Ajith et al. (2014) are used to find $x$, as a result of which the following equation can be obtained (Huerta et al., 2017):
\begin{equation}
\label{eq:pre_hyb}
 \frac{\mathrm{d}x}{\mathrm{d}t}= \tfrac{64}{5} \frac{\eta x^5}{M} \left (1 + \sum_{k=2}^{12} a_{\frac{k}{2}}x^{\frac{k}{2}} \right),
\end{equation}
% \textbf{вопрос про единицы G=c=1?}
where $\eta =\mu/M$, $\mu = M_1 M_2/M$ is the reduced mass, $a_i$ is the coefficients from the paper by Huerta et al. (2017), depending on $\eta$ and $\log(x)$.
% The value of $\eta$ will be called the symmetric mass ratio in a binary system.
Formally, the expression ~\eqref{eq:pre_hyb} looks like an approximation of 6PN.
In fact, the evolution at the convergence stage is strictly described by the equations for
the energy flux up to 3PN order, and for slightly non-circular orbits up to 3.5PN order with corrections including 6PN order.

The right-hand-side in \eqref{eq:pre_hyb} is a function
which looks as a polynomial in powers of $x$, $A(x)= a_5x^5+a_6x^6+...+ a_{11}x^{11}$,
but the coefficients $a_k$ may contain some powers of \(\log(x)\).
The equation~\eqref{eq:dxdt} will then look like
\begin{equation}
\frac{\mathrm{d}x}{\mathrm{d}t}= A(x).
\label{Ez}
\end{equation}

Here is the crucial difference between our work and that of Buskirk and
Babiuc-Hamilton~(2019).
They use \textbf{numerical} integration of the equation~\eqref{Ez}, which contradicts the
title of their article ``fully analytical form'' of the equation.
The main idea of our work is based on the following observation.
Instead of directly solving the equation~\eqref{Ez} we note that it can be rewritten as
\begin{equation}
\label{eq:Ax}
\frac{\mathrm{d}x}{A(x)}= \mathrm{d}t,
\end{equation}
using the method of separating variables.

Let us define
\begin{equation}
\label{PNparam}
 I_{A}(x) \equiv \int_{x_{\rm low}}^x \frac{\mathrm{d}\zeta}{A(\zeta)},
\end{equation}
where is the lower limit of integration --- the value of the lower bound PN of the parameter, $x_{\rm low}$ --- see below, the expression
\eqref{xlow}.

This integral cannot be solved analytically, however, it is possible to carry out transformations that will allow us to obtain an analytical result in good approximation.

To do this, we pull out the term $a_5 x^5 \equiv a x^5$, writing down
$A(x) \equiv ax^5+b(x)$, and rewrite the expression as
\begin{equation}
 I_{A}(x) = \int_{x_{\rm low}}^x \frac{\mathrm{d}\zeta}{a\zeta^5(1+\frac{b(\zeta)}{a\zeta^5})}.
\end{equation}
Now, noting that ${\zeta}\ll 1$, we apply the {\tt taylor} procedure of the {\sc Maxima} package to the following expression at $\zeta=0$:
\begin{equation}
 \frac{1}{1+\frac{b(\zeta)}{a\zeta^5}} .
\end{equation}

As is known, the {\tt taylor} procedure for non-analytic functions at zero, especially radicals
and logarithms, uses other approximations, which is not a pure Taylor expansion.
Our function $b(\zeta)$ has many logarithms and fractional powers.
As an example, we present the expansion up to the 2nd order to show the {\sc Maxima} output:
\begin{equation}
 I_{A}(x) =\int_{x_{\rm low}}^x    \frac{M}{2.599\cdot10^6 \eta \zeta^5}\bigg[1+ (2.2113+2.75  \eta )\zeta-12.566 \zeta^{1.5} +(3.0103+5.3859  \eta+4.2847  \eta^2) \zeta^2 \bigg] \mathrm{d}\zeta .
\label{powerlog}
\end{equation}
The {\sc Wolfram} procedure {\tt Series} outputs an equivalent expression.
Both {\tt taylor} and {\tt Series} can construct standard Taylor series,
as well as certain expansions involving negative powers, fractional powers, and logarithms.

Having expansions like \eqref{powerlog}, we easily obtain an analytical expression of the
integration result, the full form of which is given in the
Appendix~\ref{subsec:analyt_coeff}.
In our calculations, we used expansions up to the 4th and 6th orders and compared
their accuracy (see the tables below).

The equation~\eqref{eq:Ax} will then have the form
\begin{equation}
\label{P(x)}
I_{A}(x) = t+C.
\end{equation}

As can be seen from the equation above, the resulting dependence $x(t)$ is implicit.
The constant $C$ can be found as follows:
\begin{equation*}
I_{A}(x)\Bigr|_{t=0} = C,
\end{equation*}
and, considering that $x = x_{\rm low}$ at $t = 0$, we get
\begin{equation*}
C = I_{A}(x_{\rm low}) .
\end{equation*}

By similar reasoning, we obtain that
$I_{A}(x_{\rm high}) =t_{\rm cr} + I_{A}(x_{\rm low})$,
where $x_{\rm high}$ is the upper bound of the PN parameter, $t_{\rm cr}$---the time at
which the PN parameter grows significantly and the resulting formula is not
valid anymore.
This time approximately coincides with the time of merger obtained by other methods (note
that our formulae are correct only at the early stages of inspiral).
The result of the analytical calculation for $t=I_{A}(x)-I_{A}(x_{\rm low})$ is given in
Appendix~\ref{subsec:analyt_coeff}.

The lower limit of the PN parameter is determined by the detectable range of LIGO
detectors.
The physical limitations of the detector result from seismic, quantum and thermal noise,
which limit detection at frequencies below the threshold value of $f_{\rm low}=10$~Hz~
(LIGO Collaboration, 2015), as well as from technical limitations, due to which the
sensitivity of the detector is reduced at low frequencies.
Thus, signals from supermassive BH mergers are not recorded, since the frequency of such
events is below $10^{-4}$ Hz.

The orbital rotation velocity corresponding to $f_{\rm low}$ is (Blanchet, 2002)
\begin{equation}
{v_0=\omega_{\rm low} r=\left( G_N {M}  \omega_{\rm low} \right)^{1/3},}
\end{equation}
where $\omega_{\rm low}/\pi= f_ {\rm low}$ is the orbital angular frequency of the binary system.
The frequency of the gravitational wave is equal to twice the orbital frequency of the binary system (see, for example, Zasov, Postnov, 2011, section A.4; Maggiore, 2008, vol. 1, section 4.1; Poisson, Will, 2014, section 11.4.6).
Thus, the lower bound of the PN parameter is (Taney et al., 2016; Buskirk, Babiuc-Hamilton, 2019)
\begin{equation}
{x_{\rm low} = \frac{v_0^2}{c^2 } = \left (\pi f_ {\rm low}\frac{G_N {M}  }{c^3}\right )^{2/3}.}
%= 0.0028815 M.
\label{xlow}
\end{equation}

The upper bound of the PN parameter $x_{\rm high}$ in the Schwarzschild field is determined by the radius of the last stable orbit of the binary system (innermost stable circular orbit, ISCO): $r_{\rm ISCO} = 3R_{\rm Sch} = 6G_NM/c^2$.
In the Newtonian approximation, this radius corresponds to the orbital velocity: $v_{\rm ISCO} = \sqrt{G_NM/r_{\rm ISCO}} = c/\sqrt6$.
Therefore, in the zero order, $x^{0\rm{PN}}_{\rm ISCO} = 1/6$.
For the second order of the PN approximation, an amendment can be introduced that takes into account the dependence on the symmetric mass ratio  $\eta$~(Blanchet, 2024):
\begin{equation}
x_{\rm high}=x^{2 \rm{PN}}_{\rm ISCO} = \frac{1}{6} \left (1+\frac{7}{18}\eta \right).
\label{xhigh}
\end{equation}

Expressions for the angular phase and frequency in terms of the PN parameter are given as follows:
\begin{equation}
  \label{eq:PhiEv}
  \dot{\Phi}(x(t)) =\omega(x(t))=\frac{x(t)^{\frac{3}{2}}}{M}.
\end{equation}

The dependency $\omega(t)$ {(presented in the Appendix~\ref{subsec:omega})}, again, as in the expression~\eqref{P(x)}, is implicit.
To find the phase $\Phi(t)$, we take the derivative of $x$ on both parts of the expression~\eqref{P(x)}:
\begin{equation*}
\frac{\mathrm{d}I_{A}(x)}{\mathrm{d}x}=\frac{\mathrm{d}t}{\mathrm{d}x}
\end{equation*}
and substitute this into the integral: % to find the phase of $\Phi(t)$:
\begin{equation}
  {\Phi}(x) =\int \omega(x) \mathrm{d}t=\int \frac{\mathrm{d}t}{\mathrm{d}x}\omega(x)\mathrm{d}x=\int \frac{\mathrm{d}I_{A}(x)}{\mathrm{d}x}\omega(x)\mathrm{d}x,
\end{equation}
which gives an implicit dependence of the phase on time.
The result can be seen in the Appendix~\ref{subsec:phase}.

Next, we can determine the evolution of the distance $r(t)$ between rotating relativistic objects over time.
The distance between objects can be set in various PN approximation orders as follows:% ~\ref{subsec:r_coeff}):
\begin{equation}
\label{eq:r}
r(t)= M(r^{0\,\rm{PN}}x(t)^{-1} + r^{1\,\rm{PN}} + r^{2\,\rm{PN}}x(t)+r^{3\,\rm{PN}}x(t)^2),
\end{equation}
where $r^{i\,\rm{PN}}$ are coefficients of the corresponding order of PN approximation.
Following Buskirk and Babiuc-Hamilton~(2019), we take these coefficients from the
paper by Hinder~(2010).
Analytical formulas for $r$ and $\dot r$ are given in the Appendix~\ref{subsec:r} --- \ref{subsec:rdot}.

In a linear approximation, gravitational waves can be represented
as transverse waves with two independent polarization modes, $h_{+}$ and $h_{\times}$.

Formulae for polarizations of GW (Hinder, 2010):
\begin{equation}
\label{eq:h_plus}
h_{+}(t) =-2G_N\frac{M\eta}{Rc^4}
\bigg[
\left(-\dot r^2+r^2\dot \Phi^2+\frac{G_NM}{r}\right) \cos 2 \Phi
+ 2 r \dot r \dot \Phi \sin 2\Phi
\bigg ],
\end{equation}
\begin{equation}
\label{eq:h_cross}
h_{\times}(t) =2G_N\frac{M\eta}{Rc^4}
\bigg[\left(\dot r^2-r^2\dot \Phi^2-\frac{G_NM}{r}\right) \sin 2\Phi
- 2 r \dot r \dot \Phi \cos 2\Phi \bigg].
\end{equation}

The GW form is expressed in terms of polarization modes as follows (Apostolatos et al., 1994):
\begin{equation}
\label{eq:completeins}
h(t) =  h_{+}(t) -i  h_{\times}(t).
\end{equation}

Following Buskirk and Babiuc-Hamilton (2019), we take the angle of inclination of the orbit ~$\iota$ (the angle between the vector of the orbital angular momentum and the observation line) to be zero.
The GW radiation is maximal along the axis of the orbital moment (see Maggiore, 2008, vol. 1, section 3.6), which creates an optimal orientation for its detection.

\section{Results and Discussion}

The analytical relationship of the PN parameter $x$ and the time $t$, calculated using the
formula~\eqref{PNparam} produced by {\sc Maxima}, is given in the
Appendix~\ref{subsec:analyt_coeff}.

The tables~\ref{BHBH}~--~\ref{NSNS} indicate at what values $[x;~t]$ the discrepancy
between the analytical and numerical calculations for PN parameter $x$ reaches
$10^{-4} \ \%$ --- 1 \% for double systems with different initial masses for
a variaty of approximations (``Taylor'' series expansion near zero to orders 2, 4, and 6).
The accuracy was estimated by determining the absolute error, normalized for the critical time $t_{cr}$.
It can be seen that for many binary BH systems (see Table ~\ref{BHBH}),
the differences do not exceed 1\%, which indicates the accuracy of the analytical formula obtained.
The formulae for the phase given in appendix~\ref{subsec:phase} have approximately the same accuracy.

\begin{table}[htp]
\centering
\caption{Comparison of analytical and numerical calculations for different initial masses of BH+BH/NS (in solar masses).
$M_1$, $M_2$ --- component masses, $x_{\rm low}$ --- the initial value of the PN parameter determined by the frequency threshold when the signal enters the detection range
LIGO, $t_{\rm cr}$ --- the time at which the value of the PN parameter grows
significantly and the resulting formula stops working, [$x$; $t$] --- the value of the PN
parameter and the corresponding time when the discrepancy between analytical and numerical
calculations reaches the value indicated in the column header.}
\centerline{%
\begin{tabular}{|c|c|c|c|c|c|c|}
\hline
\multicolumn{6}{|c|}{Taylor 2nd order}\\ \hline
{\multirow{2}{*}{$M_1+M_2$}} & {\multirow{2}{*}{$x_{\rm low}$}} & {\multirow{2}{*}{$t_{\rm cr}$}} & \multicolumn{3}{c|}{$x; \ t$} \\
\cline{4-6}
\multicolumn{1}{|c|}{} & &  & $> 0.01\ \%$ & $>0.1\ \%$ & $>1\ \%$\\
\hline
  $20+20$  & 0.03370 & 12.113  &0.034; 0.182 & 0.035; 1.741& 0.058; 10.71   \\
 { $15+1.5$   } &{0.01868} & {158.76}&{0.019; 9.110} & {0.022; 75.95}& ---           \\
  $10+10$  & 0.02123 & 38.747 & 0.021; 1.815& 0.024; 15.84&  ---         \\
\hline
\multicolumn{6}{|c|}{Taylor 4th order}\\ \hline
{\multirow{2}{*}{$M_1+M_2$}} & {\multirow{2}{*}{$x_{\rm low}$}} & {\multirow{2}{*}{$t_{\rm cr}$}} & \multicolumn{3}{c|}{$x; \ t$} \\
\cline{4-6}
\multicolumn{1}{|c|}{} & &  & $> 0.001\ \%$ & $>0.01\ \%$ & $>0.1\ \%$\\
\hline
  $20+20$  & 0.03370 & 12.113  & 0.035; 1.977&0.049; 9.542 &  ---  \\
  {$15+1.5$  }  &{0.01868 }& {158.76} & {0.025; 106.7}& ---& ---           \\
  $10+10$  & 0.02123 & 38.747 & 0.029; 28.58& ---&  ---         \\
\hline
  \multicolumn{6}{|c|}{Taylor 6th order}\\ \hline
{\multirow{2}{*}{$M_1+M_2$}} & {\multirow{2}{*}{$x_{\rm low}$}} & {\multirow{2}{*}{$t_{\rm cr}$}} & \multicolumn{3}{c|}{$x; \ t$} \\
\cline{4-6}

\multicolumn{1}{|c|}{} & &&$> 0.001\ \%$& $> 0.01\ \%$ & $>0.1\ \%$ \\
\hline
  $20 + 20$ &  0.03370   & 11.856 &0.048; 9.081& 0.094; 11.77 & ---     \\
  { $15+1.5$}    & {0.01868} & {158.76 }&{0.066; 158.0}& ---  & ---             \\
  $10 + 10$ &  0.02123   & 38.511 &0.079; 38.38& --- & ---    \\

\hline
\end{tabular}
}
\label{BHBH}
\end{table}

For binary NS systems (see table~\ref{NSNS}), the formula error does not exceed fractions of a percent.
This is due to the increased time before merging: the formula works fine for small values of the PN parameter, which increases sharply only at the moment of merging.
Expansion to the sixth order gives the highest accuracy, on the order of a thousandth of a percent.

\begin{table}[htp]
\centering
\caption{Comparison of analytical and numerical calculations for different initial masses of NS+NS (in solar masses).
$M_1$, $M_2$ --- component masses, $x_{\rm low}$ --- the initial value of the PN parameter determined by the frequency threshold when the signal enters the detection range
LIGO, $t_{\rm cr}$ --- the time at which the value of the PN parameter increases
significantly and the resulting formula stops working, [$x$; $t$] --- the value of the PN
parameter and the corresponding time when the discrepancy between analytical and numerical
calculations reaches the value indicated in the column header.}
\centerline{%
\begin{tabular}{|c|c|c|c|c|c|c|}
\hline
\multicolumn{6}{|c|}{Taylor 2nd order}\\ \hline
{\multirow{2}{*}{$M_1+M_2$}} & {\multirow{2}{*}{$x_{\rm low}$}} & {\multirow{2}{*}{$t_{\rm cr}$}} & \multicolumn{3}{c|}{$x; \ t$} \\
\cline{4-6}
\multicolumn{1}{|c|}{} & &  & $>10^{-4}\ \% $ & $>0.001\ \%$ & $>0.01\ \%$\\
\hline
  $2 + 2$  &0.007260 & 562.107 & 0.007; 3.700 & 0.007; 36.90   &0.009; 303.20    \\
  $1.5+1.5$& 0.005993& 906.189 & 0.006; 9.104& 0.006; 94.30   &0.008; 679.0    \\
  $1 + 1$  &0.004574 & 1776.75 & 0.005; 37.11 & 0.005; 349.2  & 0.013; 1750  \\
\hline
\multicolumn{6}{|c|}{Taylor 4th order}\\ \hline
{\multirow{2}{*}{$M_1+M_2$}} & {\multirow{2}{*}{$x_{\rm low}$}} & {\multirow{2}{*}{$t_{\rm cr}$}} & \multicolumn{3}{c|}{$x; \ t$} \\
\cline{4-6}
\multicolumn{1}{|c|}{} & &  & $>10^{-5}\ \% $ & $>10^{-4}\ \% $& $>0.001\ \%$\\
\hline
  $2 + 2$  &0.007260 & 561.886 & 0.009; 403.3 & 0.066; 561.82   &---    \\
  $1.5+1.5$& 0.005993& 906.189 & 0.012; 846.2& 0.028; 903.9   &---    \\
  $1 + 1$  &0.004574 & 1776.75 & 0.024; 1773.7 & 0.027; 1774.8  & ---  \\
\hline
\multicolumn{6}{|c|}{Taylor 6th order}\\ \hline
{\multirow{2}{*}{$M_1+M_2$}} & {\multirow{2}{*}{$x_{\rm low}$}} & {\multirow{2}{*}{$t_{\rm cr}$}} & \multicolumn{3}{c|}{$x; \ t$} \\
\cline{4-6}
\multicolumn{1}{|c|}{} & &  &$>10^{-5}\ \% $& $>10^{-4}\ \% $ & $>0.001\ \%$  \\
\hline
  $2 + 2$   &  0.007260  & 561.884 &0.031; 559.5&0.063; 561.80 &---        \\
  $1.5+1.5$ & 0.005993   & 905.966 &0.027; 903.4&0.062; 905.90  &---   \\
  $1 + 1$   &  0.004574  & 1776.52 &0.043; 1776.0&---  &---     \\
\hline
\end{tabular}
}
\label{NSNS}
\end{table}

The obtained analytical formula can facilitate and accelerate the detection of merging
objects at the earliest stages.
Analytical formulae explicitly relate the observed signal parameters
(frequency, phase, amplitude) to the physical characteristics of the system
(masses, orbital parameters) and this significantly increases the detection time of the signal.

This is especially true for NS systems.
For example, for the NS~+~NS system with masses of $1.5 M_{\odot}$ (standard mass of NS;
see table~\ref{NSNS}) the time to merging (the time for which the system can be monitored)
reaches 900 seconds.
For a similar GW170817 event, the detection time was about 100 seconds~(Abbott et al., 2017).

\begin{figure}[H]
\centering
\includegraphics[width=0.63\textwidth]{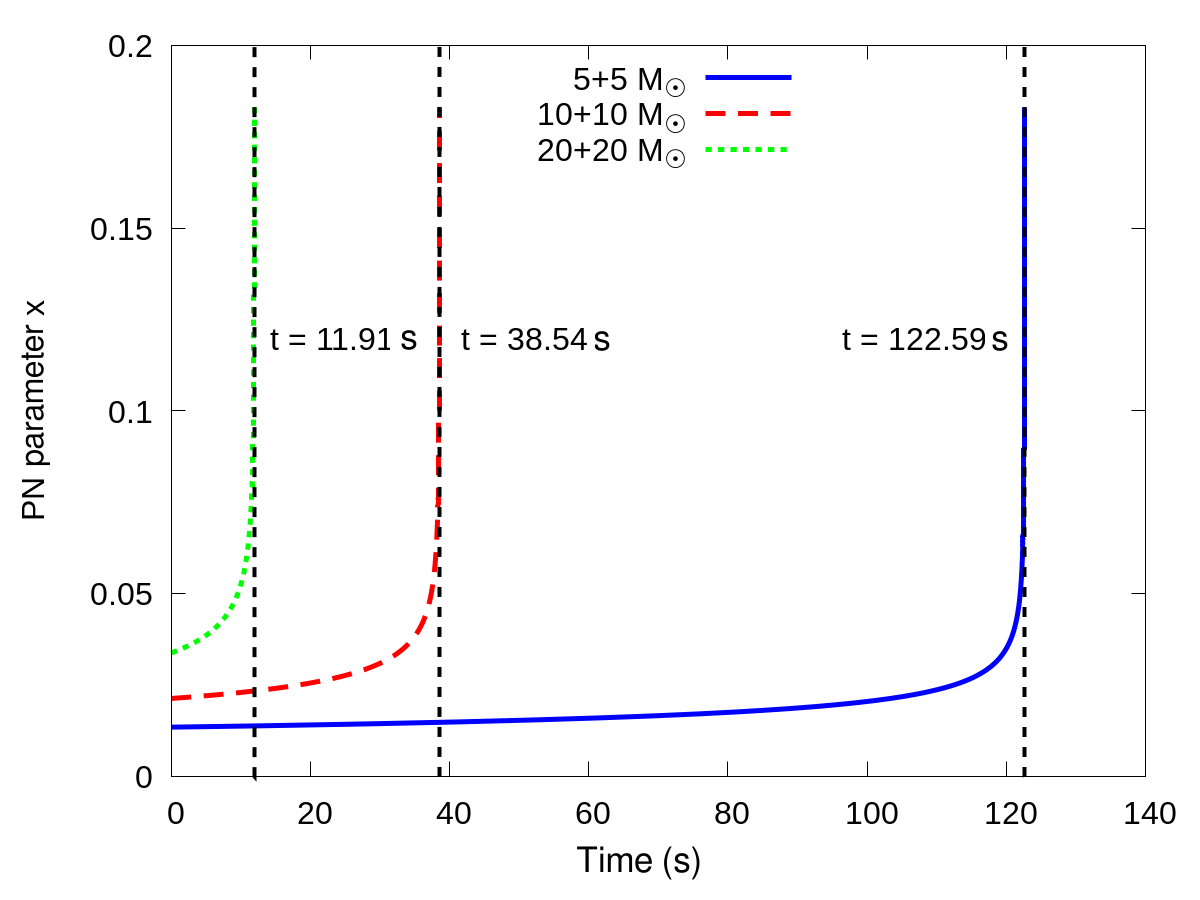}
%\epsfysize=8cm
%\centering{\epsfbox{f10.ai}}
\caption{ Evolution of the PN parameter over time for various BH systems}
 \label{pnbh}
\end{figure}

\begin{figure}[H]
\centering
\includegraphics[width=0.63\textwidth]{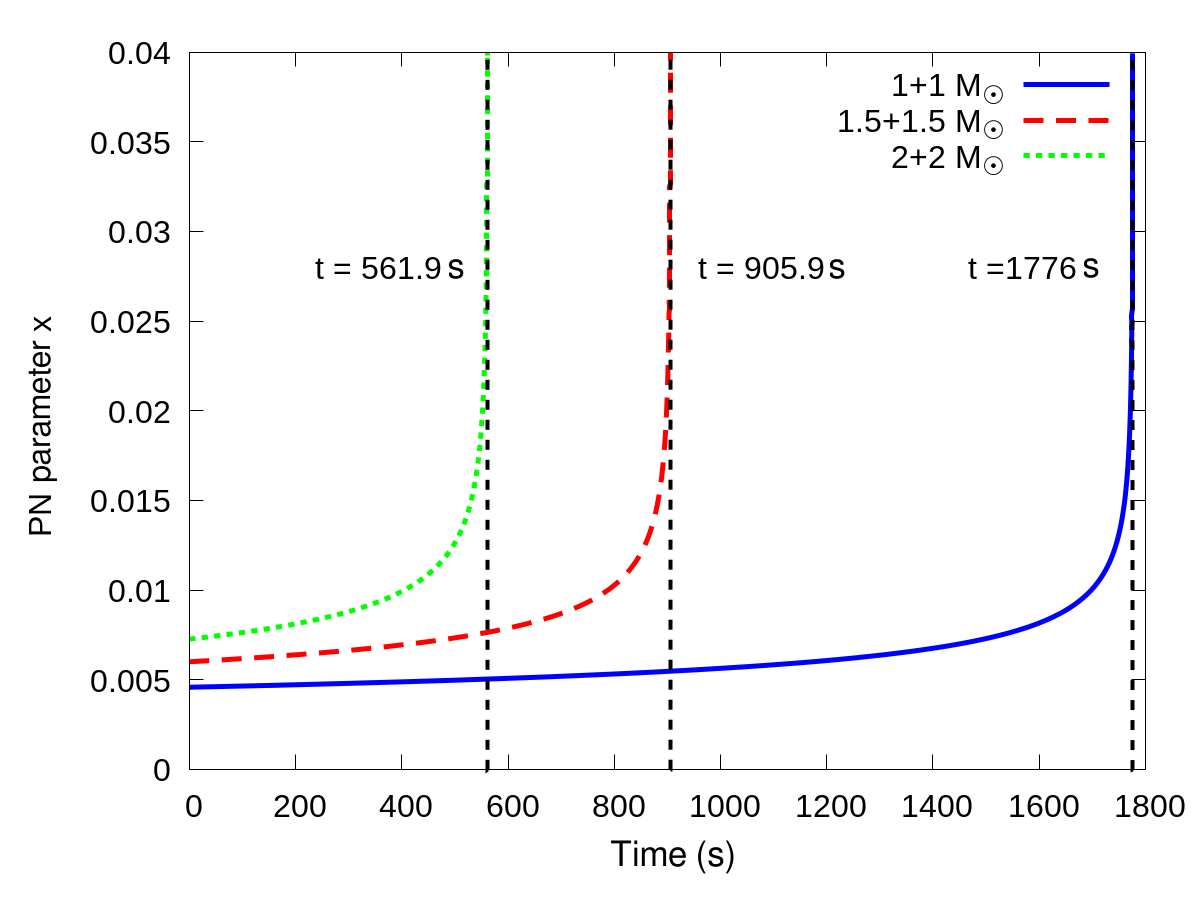}
%\epsfysize=8cm
%\centering{\epsfbox{f10.ai}}
\caption{ Evolution of the PN parameter over time for various NS systems}
 \label{pn-ns}
\end{figure}

In the Figures ~\ref{pnbh} -- \ref{pn-ns}, one can see the evolution of the PN parameter
over time for different BH {and NS} systems.
The time it takes to start monitoring depends on the mass of the system and ranges from
tens of seconds (for massive binary systems) to hundreds (for less massive ones) for
different systems.
As we approach the moment of merging, the PN parameter grows dramatically, and the
resulting formula becomes inapplicable.

\begin{figure}[H]
\centering
\includegraphics[width=0.56\textwidth]{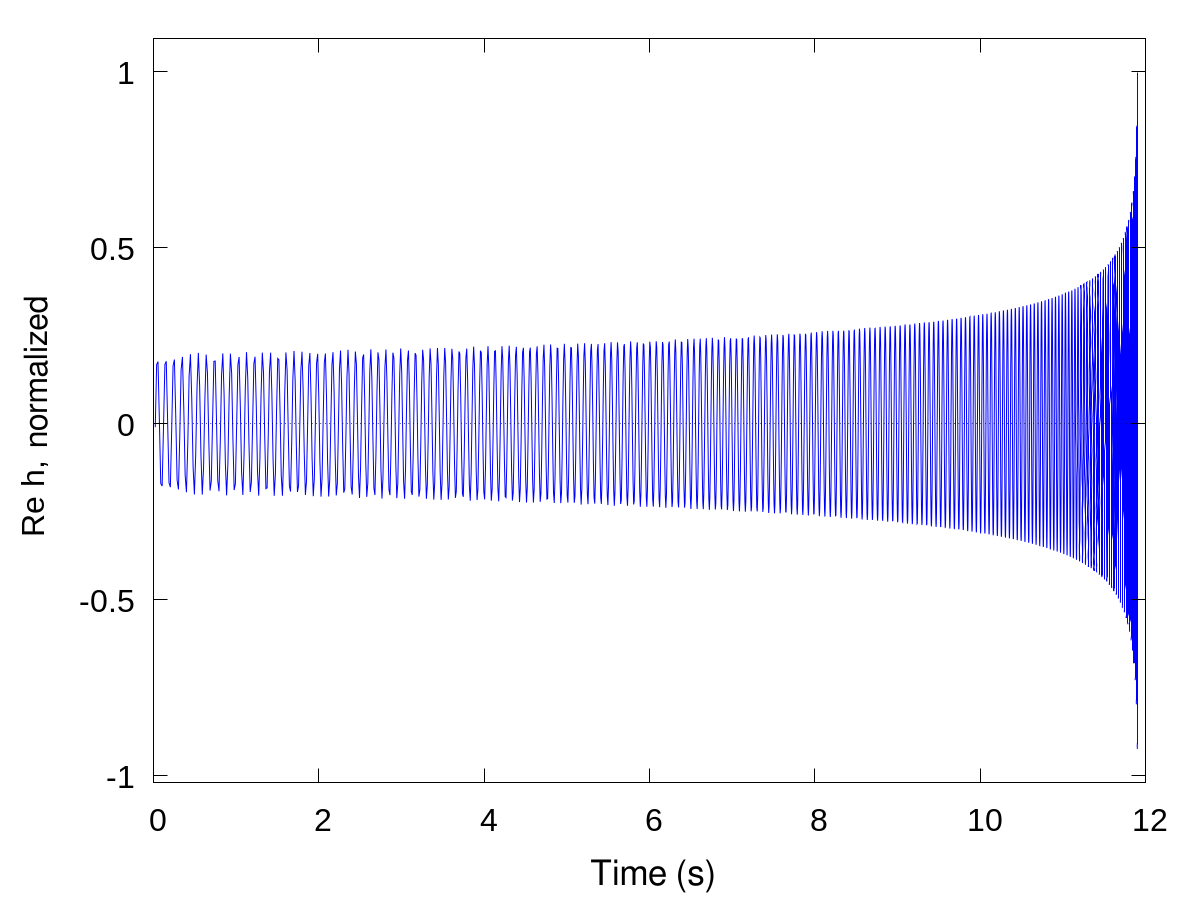}
%\epsfysize=8cm
%\centering{\epsfbox{f10.ai}}
\caption{ GW form for the BH system~$ (20+20) M_\odot$ }
 \label{renormalized}
\end{figure}

 The final result is the analytical template (form) of the GW, shown in the Figure~\ref{renormalized}.

\begin{figure}[H]
\centering
\includegraphics[width=0.56\textwidth]{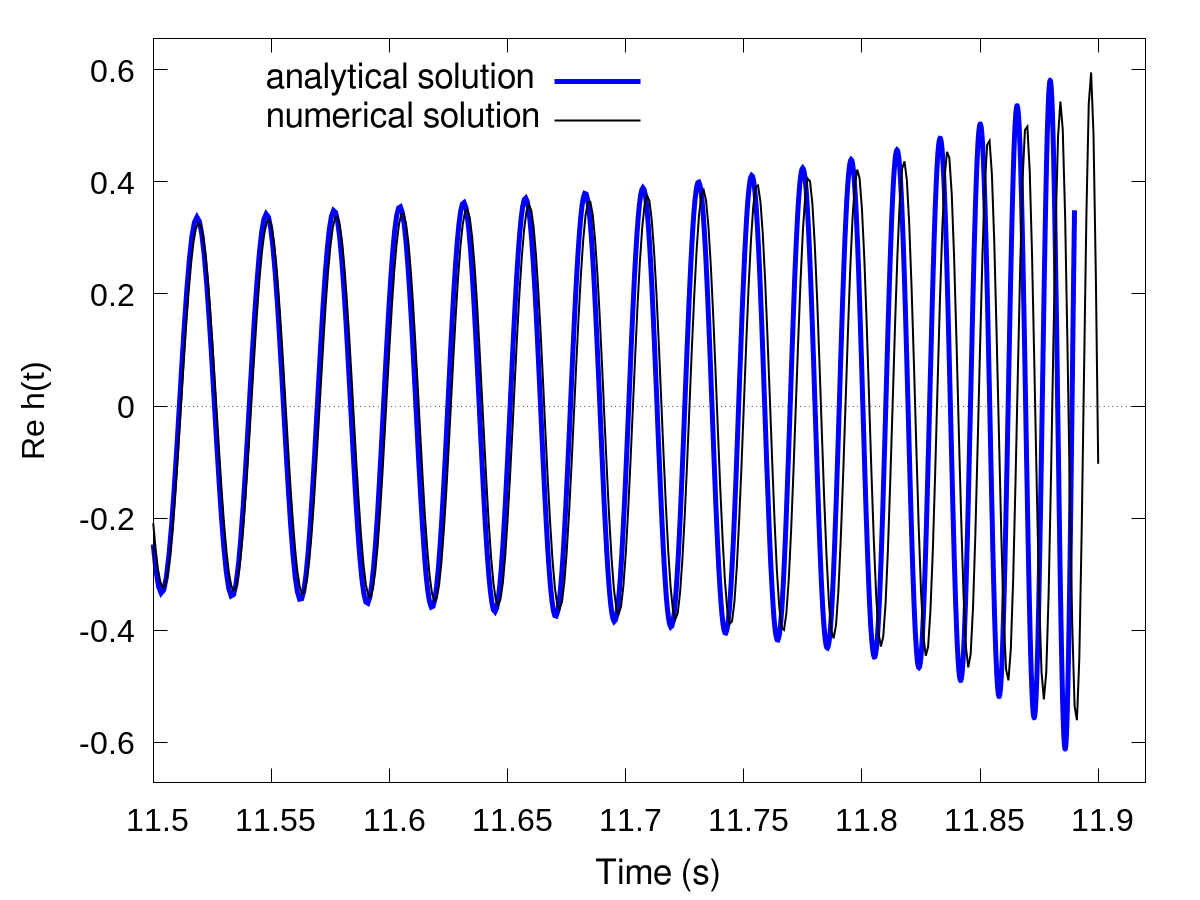}
%\epsfysize=8cm
%\centering{\epsfbox{f10.ai}}
\caption{ Comparison of the numerical and analytical form of the GW for the BH system ~$(20+20)M_\odot$ half a second before the merger}
 \label{waveformcomparison2}
\end{figure}

\begin{figure}[H]
\centering
\includegraphics[width=0.65\textwidth]{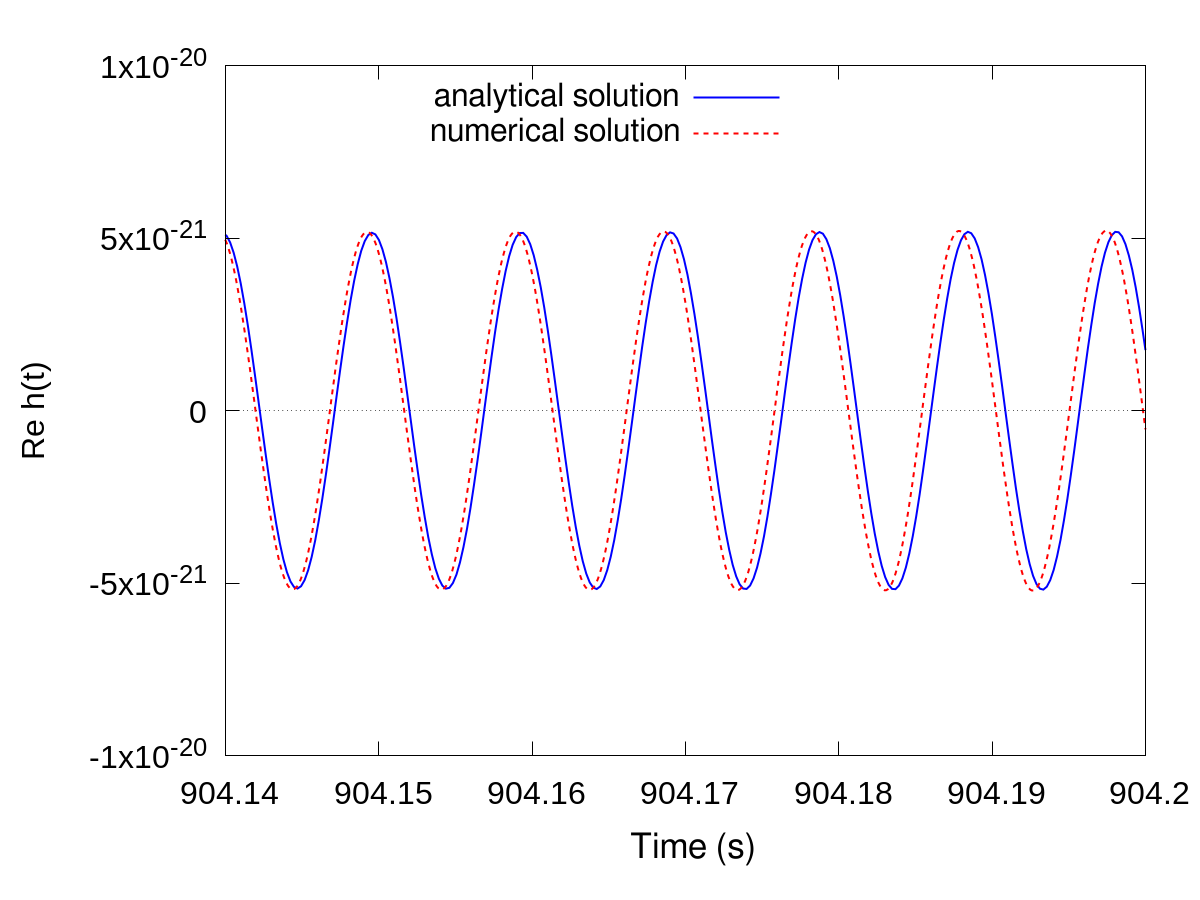}
%\epsfysize=8cm
%\centering{\epsfbox{f10.ai}}
\caption{Comparison of the numerical and analytical form of the GW in the last seconds before the merger for the system $(1.5+1.5)M\odot$ (for this system, the critical time is $t_{cr}=905.9$~s --- see Fig.~\ref{pn-ns})}
 \label{1.5NS}
\end{figure}

The Figures~\ref{waveformcomparison2}---\ref{1.5NS} show a comparison of the GW-form calculated numerically using {\sc Mathematica} and analytically in {\sc Maxima}.
The figure~\ref{waveformcomparison2} shows the shape of the GW for the BH system.
The characteristic sinusoidal shape of the GW is clearly visible here in the last second just before the merger, when the amplitude increases noticeably.
The analytical formula and numerical calculation begin to diverge only at the last second before the merger.
The figure \ref{1.5NS} shows the last milliseconds before merging for the NS system~$(1.5+1.5)M\odot$.

%%%%%%%%%%%%%%%%%%%%%%%%%%%%%%%%
%%%%%%%%%%%%%%%%%%%%%%%%%%%%%%%%

\begin{figure}[H]
\centering
\includegraphics[width=0.7\textwidth]{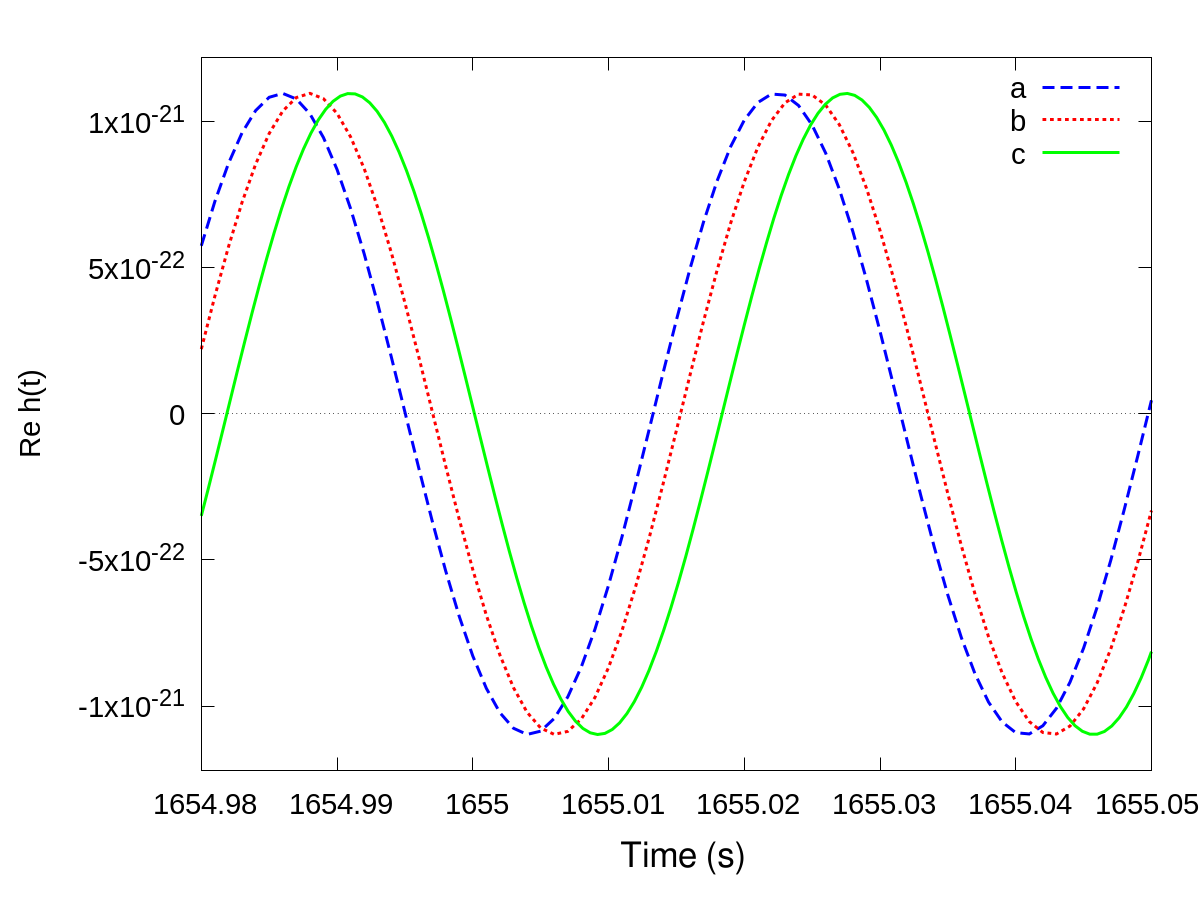}
%\epsfysize=8cm
%\centering{\epsfbox{f10.ai}}
\caption{ A comparison of the analytically obtained form of GW and numerical expressions calculated with different accuracy for the NZ system $(1+1)M\odot$ (for this system, the critical time is $t_{cr}=1776$~s --- see Fig.~\ref{pn-ns}). a --- numerical calculation (accuracy is set by default in {\sc Mathematica}), b --- numerical calculation (accuracy 8), c --- analytical calculation}
 \label{wolframaccuracy}
\end{figure}

In the tables and graphs presented above, we checked the accuracy of the analytical formula by comparing it with the result of a numerical expression calculated using {\sc Mathematica}.
But it is also interesting to check the accuracy of the numerical method itself {\sc} used in {\sc Mathematica}, since Buskirk and Babiuc-Hamilton~(2019) did not provide an indication of the accepted accuracy of the numerical method.
The figure \ref{wolframaccuracy} shows a comparison of our analytical formula and two numerical results.
{\sc NDSolve} for different accuracy for the NS system~$(1+1) M\odot$.
Obviously, the amplitudes and periods are reproduced by different methods with comparable accuracy, but differences in phase irreversibly arise due to the large number of periods of the shape of the GW.

As is well known, the so-called chirp mass plays an important role in calculating the phase of GW (Cutler and Flanagan, 1994):
\begin{equation}
M_{\rm{chirp}}=\frac{(M_1M_2)^{3/5}}{M^{1/5}}
\label{chirp}
\end{equation}

At the same time, the chirp frequency describes the rate at which this frequency increases
during the convergence stage, and is primarily determined by the chirp mass of the binary
system.
This dependence is expressed by the equation:
{\begin{equation}
\frac{\mathrm{d} \omega}{\mathrm{d}t} =\bigg(\frac{96}{5}\frac{G_N M_{\rm{chirp}} }{c^3}\bigg)^{5/3} \omega^{11/3}
\label{deriv_omega}
\end{equation}}

The chirp mass is not explicitly included in the formulas we use, however, it is interesting to compare the chirp mass values using the formulas~\ref{chirp} and ~\ref{deriv_omega}.
The Table~\ref{chirpmass} shows a comparison of the chirp mass values calculated for the neutron star system.

\begin{table}[htp]
  \caption{{Comparison of the accuracy of the mass chirp for the 2 +1 neutron star system $M_{\odot}$ for different values of the PN parameter}}
\begin{center}
\begin{tabular}{|c|c|c|c|c|c|}
\hline
{NS + NS}&{0PN} &  {0.0001$ \cdot x_{\rm low}$}& {0.01$ \cdot x_{\rm low}$ }& {$ x_{\rm low}$} & {$2 \cdot x_{\rm low}$}  \\
\hline
 {2 + 1}&  {1.216729}&  {1.216727}& { 1.216609}&  {1.208659}& {1.204167}\\
 \hline
\end{tabular}
\end{center}
\label{chirpmass}
\end{table}

It can be seen that at values of $x$ 4 orders of magnitude lower than the
threshold of the PN parameter $x_{\rm low}$, the values of
the chirp masses perfectly match, since the formula~\ref{chirp} is correct in the 0PN
approximation.
However, already at $x = x_{\rm low}$, a noticeable error appears, which grows with the growth of $x$.

 So, thanks to the new analytical formula, it will be possible to detect gravitational wave events much earlier.
Due to this, high calculation speed and prompt event detection are achieved.
The explicit dependence on the parameters (mass, phase, distance between objects) is convenient for subsequent data analysis.

 \section*{\centering{Results}}
% \addcontentsline{toc}{section}{ЗАКЛЮЧЕНИЕ}
%

A fully analytical template for gravitational waveforms for close binary systems at the early stage of merging is obtained.

The use of analytical templates of gravitational waves plays a key role in
gravitational-wave astronomy, as they make it possible to quickly process data from detectors and
effectively identify significant events.
Comparing the observed signals with theoretical models significantly speeds up their
detection.
That is why the creation of accurate analytical and numerical models describing this
process remains one of the most important tasks of modern theoretical physics.

The formula obtained in this article may help one in quick detection and study of the
gravitational wave signal, especially in systems with low masses.
For example, for neutron star binaries, it is possible to detect and monitor them at
the earliest stages of coalescence, up to thousands of seconds before ringdown.

We compare the numerical calculations with our analytical formula, evaluate its accuracy and limits of applicability.
We find that the error of the formula does not exceed 1\% for all the studied pairs of
relativistic objects (BH~+~BH, BH~+~NS, and NS~+~NS), and for NS~+~NS binaries
the error is less than $10^{-4}\ \%$ for most of the interval until the critical time $t_{cr}$.

We thank V.E.~Valiulin, N.K.~Poraiko, A.V.~Yudin, and N.I.~Kramarev for their interest in our
studies and useful discussions.

{We are grateful to the referees of the Astronomy Letters for their valuable comments,
which helped us to significantly improve the presentation of our results.}

S.B. would like to acknowledge the support of Russian Science Foundation
in the search of effective approximations for gravitational waveforms (RSF project 25-22-00293).
\setcounter{subsection}{0}

\begin{appendices}
\renewcommand{\thesection}{\Alph{section}}\renewcommand{\sectionmark}[1]{}
\renewcommand{\thesubsection}{\thesection.\arabic{subsection}}
\renewcommand{\thesubsubsection}{\thesubsection.\arabic{subsubsection}}

\section{Appendix}
The final formulae are listed on GitHub.
Link to the repository: \href{https://github.com/sblinnikov/gw-forms-analytics}{\textcolor{RoyalBlue}{https://github.com/sblinnikov/gw-forms-analytics}}.
\subsection{Analytical expression for PN parameter}
\label{subsec:analyt_coeff}
The analytical result obtained as a result of integration, depending on such parameters as $x$ (PN parameter~\ref{PNparameter}), $M$ (total mass of the system), $\eta$ (symmetric mass ratio) are given below:

%-------------------------------------------------------%

\begin{equation*}
\begin{gathered}
\operatorname{t}(\ensuremath{x}{, \ }M{, \ }\eta)  \eta  x^4  M^{5/3}=
\begin{minipage}[t]{.68\displaywidth}
  \raggedright\linespread{1.2}\selectfont
\begin{math}
 c_1  (-1.50413 \cdot 10^{-2} + 4.67964 \cdot 10^{-5}  M +  c_3 (-0.49049  - 1.90751 \cdot 10^{-3}  M)
+ 6.90126 \cdot 10^{-11}  M^{8/3} \log(2.88225 \cdot 10^{-3}
M^{2/3})^2 \cdot (0.65916 + 8.30459 \cdot 10^{-6}  M^{4/3} + 5.88205  \eta))
+  c_2  (-1.50413 \cdot 10^{-2} + 0.30242    x^{3/2}
+  c_4     (-0.49049  - 12.3273 x^{3/2})
+ x^4  \log(x)^2  (0.65916  + 0.99966   x^2 + 5.88205 \eta))
+(1.44930 - 0.20476  \eta) \cdot ( 4.45995 \cdot 10^{-7}  c_1 M^{5/3}+ c_2  x^{5/2})
+(-4.43479 \cdot 10^{-2} - 5.51514 \cdot 10^{-2} \eta) \cdot (  2.88225 \cdot 10^{-3}  c_1 M^{2/3}+ c_2  x)
+(5.73679 + 4.73180 \eta - 1.73876 \eta^2) \cdot ( 1.28547 \cdot 10^{-9}  c_1 M^{7/3}+ c_2  x^{7/2})
+(-9.05580 \cdot 10^{-2} - 0.16202  \eta - 0.12889  \eta^2) \cdot ( 8.30738 \cdot 10 ^{-6}  c_1 M^{4/3}+ c_2  x^2)
+(77.9184 - 232.665  \eta - 160.354  \eta^2 - 35.6463  \eta^3) \cdot ( 3.70505 \cdot 10^{-12}  c_1 M^3+ c_2  x^{9/2})
+(-1.81803 - 9.95228 \eta + 0.13240  \eta^2 - 0.29671  \eta^3) \cdot (  2.39440 \cdot 10^{-8 }c_1 M^2+ c_2  x^3)
+(230.182  - 43.6834  \eta - 258.419  \eta^2 - 314.396  \eta^3 - 58.5971  \eta^4) \cdot ( 1.06789 \cdot 10^{-14}  c_1 M^{11/3}+ c_2  x^{11/2})
+(-133.656  + 154.205  \eta + 88.0425  \eta^2 + 65.8957  \eta^3 + 5.90339 \eta^4 + 0.50839  \eta^5) \cdot ( 1.98912 \cdot 10^{-13}  c_1 M^{10/3}+ c_2  x^5)
+(-441.966  + 972.052  \eta + 1122.42  \eta^2 + 227.437  \eta^3 + 64.3629  \eta^4 + 10.4714  \eta^5 + 0.5  \eta^6) \cdot ( 5.73314 \cdot 10^{-16} c_1 M^4+ c_2  x^6)
+(-12.7081 - 72.6159  \eta) \cdot ( 4.45995 \cdot 10^{-7} c_1 M^{5/3}   c_3+ c_2  x^{5/2}   c_4)
+(1.76923 - 14.6850  \eta - 11.1794  \eta^2) \cdot ( 8.30738 \cdot 10 ^{-6}  c_1 M^{4/3}   c_3+ c_2  x^2   c_4)
+(4.88255 - 215.061  \eta - 149.023  \eta^2 - 35.7480  \eta^3) \cdot (  2.39440 \cdot 10^{-8}  c_1 M^2   c_3+ c_2  x^3   c_4)
+(-13.4691 + 7.63569 \eta + 28.8131  \eta^2 + 0.28618  \eta^3 + 0.33024  \eta^4) \cdot ( 2.88225 \cdot 10^{-3}  c_1 M^{2/3}   c_3+ c_2    c_4 x)
\end{math}
  \end{minipage}
\end{gathered}
\end{equation*}

\begin{flalign*}
c_1 &= (-9.26755) \cdot 10^4  x^4 \\
c_2 &= 6.39578 \cdot 10^{-6}  M^{8/3}\\
c_3 &= 2.39440 \cdot 10^{-8}  \cdot \log(2.88225 \cdot 10^{-3} \cdot M^{2/3}) M^2\\
c_4 &= \log(x)x^3
\end{flalign*}

\pagebreak
\subsection{Analytical template for gravitational wave}
\label{AnalytFormula}

The final formula of the analytical template of a gravitational wave, depending on the parameters $x$ (PN parameter~\ref{PNparameter}), $M$ (total mass of the system), $\eta$ (symmetric mass ratio) has the following form:
\begin{flalign*}
    \operatorname{Re(h)} (\ensuremath{x}{, \ }M{, \ }\eta) & \operatorname{=} -\frac{2 G M\, \eta }{Rc^4} \bigg[\bigg(-{{ \dot{\mathrm{r}}^{2}\left( \ensuremath{x}{, \ }M{, \ }\eta \right) }}+{{ \operatorname{r}^{2}\left( \ensuremath{x}{, \ }M{, \ }\eta \right) }}    {{{\omega}^{2}\left( \ensuremath{x}{, \ }M \right) }}+\frac{M}{\operatorname{r}\left( \ensuremath{x}{, \ }M{, \ }\eta \right) } \bigg)\\&\times \cos{\left( 2 \operatorname{\Phi}\left( \ensuremath{x}{, \ }M{, \ }\eta \right) \right) }+2 \operatorname{r}\left( \ensuremath{x}{, \ }M{, \ }\eta \right) \dot{\mathrm{r}}\left( \ensuremath{x}{, \ }M{, \ }\eta \right) {\omega}\left( \ensuremath{x}{, \ }M\right)  \sin{\left( 2 \operatorname{\Phi}\left( \ensuremath{x}{, \ }M{, \ }\eta \right) \right) }\bigg]
\end{flalign*}
 \begin{flalign*}
 \hspace{1.25cm}\text{where }&Re(h) \text{ --- real part } h(t)\text{;}&&\\
 &R\text{ --- distance from the double to the detector}
 \end{flalign*}

and auxiliary functions:

%-------------------------------------------------------%

 \subsubsection{The orbital frequency function}
 \label{subsec:omega}
{
\vspace{-0.5cm}
\begin{flalign*}
    \hspace{1.25cm}&{{\omega}}\left( \ensuremath{{x}}{,}M\right) \operatorname{=}\frac{2.03026\cdot10^{5}x^{3/2}}{M}&&
\end{flalign*}
}

\pagebreak
 \subsubsection{Phase function}
 \label{subsec:phase}
{
\begin{equation*}
\begin{gathered}
\operatorname{\Phi}(\ensuremath{x}{, \ }M{, \ }\eta)  \eta M^{5/3} x^{5/2}=
\begin{minipage}[t]{.68\displaywidth}
  \raggedright\linespread{1.2}\selectfont
\begin{math}
p_1    (-0.03125+ 1.51914\cdot 10^{-4}M
	+ \log(p_2)     (3.05001\cdot 10^{-8}M^2 - 1.48268\cdot 10^{-11}M^3)
	+ 6.90126\cdot 10^{-11}M^{8/3} \log(p_2)^2(4.58586\cdot 10^{-16} + 6.16205\cdot 10^{-6}M^{4/3} + 3.66869\cdot 10^{-15}\eta ))
+ M^{5/3}     (-0.03125 + 0.98175x^{3/2}
	+ \log(x)     (1.27381 x^3 - 4.00179 x^{9/2})
	+ x^4 \log(x)^2(4.58586\cdot 10^{-16} + 0.74176x^2 + 3.66869\cdot 10^{-15}\eta ))
+(-0.11517 - 0.14323\eta )\cdot(p_1 p_2 +M^{5/3} x)
+(-0.47036- 0.84155\eta  - 0.66949\eta ^2)\cdot(8.30738\cdot 10^{-6}M^{4/3}p_1+M^{5/3}  x^2)
+(-3.72464 - 3.07214\eta  + 1.12890\eta ^2)\cdot(1.28547\cdot 10^{-9}M^{7/3}p_1+M^{5/3} x^{7/2})
+(19.2918 - 75.5294\eta  - 52.0552\eta ^2 - 11.5718\eta ^3)\cdot(3.70505\cdot 10^{-12}M^3 p_1+M^{5/3} x^{9/2})
+(0.90004+ 25.8463\eta  - 0.34385\eta ^2 + 0.77055\eta ^3)\cdot(2.39440\cdot 10^{-8}M^2 p_1+M^{5/3}  x^3)
+(146.696 - 44.0771\eta  - 167.780\eta ^2 - 204.123\eta ^3 - 38.0444\eta ^4)\cdot(1.06789\cdot 10^{-14}M^{11/3}p_1+M^{5/3} x^{11/2})
+(-12.4206 - 0.17923\eta  + 24.9427\eta ^2 + 0.24774\eta ^3 + 0.28588\eta ^4)\cdot(6.90126\cdot 10^{-11}M^{8/3}p_1+ M^{5/3} x^4)
+(-68.8702 + 75.5183\eta  + 42.2456\eta ^2 + 34.2265\eta ^3 + 3.06624\eta ^4 + 0.26406\eta ^5)\cdot(1.98912\cdot 10^{-13}M^{10/3}p_1+ M^{5/3} x^5)
+(-327.256 + 687.073\eta  + 809.148\eta ^2 + 163.076\eta ^3 + 47.7576\eta ^4 + 7.76980\eta ^5 + 0.37100\eta ^6)\cdot(5.73314\cdot 10^{-16}M^4 p_1+M^{5/3} x^6)
+(-8.25082 - 47.1462\eta )\cdot(1.06789\cdot 10^{-14}M^{11/3}p_1+M^{5/3} x^{11/2})
+(-2.82289 + 0.39884\eta )\cdot(4.45995\cdot 10^{-7} M^{5/3} p_1 \log(p_2)+M^{5/3} x^{5/2} \log(x))
+(0.91894- 7.62745\eta  - 5.80663\eta ^2)\cdot(1.98912\cdot 10^{-13}M^{10/3}p_1 \log(p_2)+M^{5/3}  x^5 \log(x))
+(3.94077 - 159.577\eta  - 110.576\eta ^2 - 26.5252\eta ^3)\cdot(5.73314\cdot 10^{-16}M^4 p_1 \log(p_2)+M^{5/3} x^6 \log(x))
+(1.14125 + 10.1839\eta  + 1.46748\cdot 10^{-14}\eta ^2 + 1.14646\cdot 10^{-16}\eta ^3)\cdot(6.90126\cdot 10^{-11}M^{8/3} p_1 \log(p_2)+M^{5/3} x^4 \log(x))
\end{math}
  \end{minipage}
\end{gathered}
\end{equation*}
}
{
\begin{flalign*}
p_1&=-2.24218\cdot10^6x^{5/2}\\
p_2&=2.88225\cdot10^{-3}M^{2/3}
\end{flalign*}
}

\pagebreak
 \subsubsection{The function of the distance between the stars}
 \label{subsec:r}

\begin{flalign*}
  \operatorname{r}\left( \ensuremath{{x}}{,}M{,}\eta \right) &= 4.92548\cdot 10^{-6}M\bigg(x^{-1} -1 +\frac{\eta}{3}  + \frac{x(  342 + 8\eta^2)}{72}
  +x^2\bigg(7.51822\eta - \frac{37\eta^2}{12}\\
  &+ \frac{2\eta^3}{81}\bigg)\bigg)
\end{flalign*}
%-------------------------------------------------------%

%%%%%%%%%%%%%%%%

 \subsubsection{The function of the derivative of the distance between the stars in time}
 \label{subsec:rdot}
{
\begin{equation*}
\begin{gathered}
\operatorname{\dot r}(\ensuremath{x}{, \ }M{, \ }\eta)  =
\begin{minipage}[t]{.68\displaywidth}
  \raggedright\linespread{1.2}\selectfont
\begin{math}
(-0.77011 x^3 \eta + 3.65804 x^5 \eta^2 - 11.5798 x^6 \eta^2 + 8.55682 \cdot 10^{-2} x^5 \eta^3 - 4.74903 x^6 \eta^3 + 3.80303 \cdot 10^{-2} x^6 \eta^4)/(6.01651\cdot 10^{-2} + 0.13304 x - 0.75606 x^{3/2} + 0.18112 x^2 - 2.17395 x^{5/2} + 1.32754 x^3 - 2.86840 x^{7/2} - 13.4691 x^4 + 26.6319 x^{9/2} - 131.887 x^5 + 332.565 x^{11/2} - 879.050 x^6 + 0.16545 x \eta + 0.32404 x^2 \eta + 0.30715 x^{5/2} \eta + 9.95228 x^3 \eta - 2.36590 x^{7/2} \eta + 7.63569 x^4 \eta - 116.332 x^{9/2} \eta + 139.520 x^5 \eta - 138.141 x^{11/2} \eta + 1729.04 x^6 \eta + 0.25779 x^2 \eta^2 - 0.13240 x^3 \eta^2 + 0.86938 x^{7/2} \eta^2 + 28.8131 x^4 \eta^2 - 80.1768 x^{9/2} \eta^2 + 76.8631 x^5 \eta^2 - 387.629 x^{11/2} \eta^2 + 2095.82 x^6 \eta^2 + 0.29671 x^3 \eta^3 + 0.28618 x^4 \eta^3 - 17.8231 x^{9/2} \eta^3 + 65.8957 x^5 \eta^3 - 471.594 x^{11/2} \eta^3 + 419.126 x^6 \eta^3 +0. 33024 x^4 \eta^4 + 5.90339 x^5 \eta^4 - 87.8956 x^{11/2} \eta^4 + 128.726 x^6 \eta^4 + 0.50840 x^5 \eta^5 + 20.9427 x^6 \eta^5 + x^6 \eta^6 +
 \log(x) \cdot (0.49049 x^3 + 1.31833 x^4  - 6.16367 x^{9/2}  + 1.76923 x^5  - 19.0622 x^{11/2}  + 11.7644 x^6  + 11.7641 x^4 \eta  - 14.6850 x^5 \eta  - 108.924 x^{11/2} \eta  - 430.123 x^6 \eta  + 1.69518\cdot 10^{-14} x^4 \eta^2  - 11.1794 x^5 \eta^2  - 298.047 x^6 \eta^2  + 1.32436\cdot 10^{-16} x^4 \eta^3  - 71.4960 x^6 \eta^3 ) + \log(x)^2 \cdot( 5.29745\cdot 10^{-16} x^4 + 1.99933 x^6 + 4.23796\cdot 10^{-15} x^4 \eta))
\end{math}
  \end{minipage}
\end{gathered}
\end{equation*}
}

\end{appendices}
\end{document}